\documentclass[conference]{IEEEtran}

\usepackage{ifpdf}

\usepackage{booktabs}
\usepackage{amssymb}
\usepackage{subcaption}
\usepackage{threeparttable}
\usepackage{url}
\usepackage{multirow}
\usepackage{makecell}

\usepackage{cite}
\usepackage[hidelinks]{hyperref}

\ifCLASSINFOpdf
  \usepackage[pdftex]{graphicx}
\else
\fi

\usepackage[font=small]{caption}

\begin{document}
%
\newif\ifanonymous
\anonymousfalse
\title{TransDot: An Area-efficient Reconfigurable Floating-Point Unit for Trans-Precision Dot-Product Accumulation for FPGA AI Engines}

\ifanonymous
\author{
}
\else
\author{
  \IEEEauthorblockN{
    Jiayi Wang\IEEEauthorrefmark{1},
    Maohua Nie\IEEEauthorrefmark{1},
    Sin-Chen Lin,
    C.-J. Richard Shi,
    Ang Li
  }
  \IEEEauthorblockA{Department of Electrical and Computer Engineering, University of Washington\\
  Email: \{jwang710, mhnie, sinchl, cjshi, angliz\}@uw.edu}
  \IEEEauthorblockA{\IEEEauthorrefmark{1}Equal contribution}
}
\fi

\maketitle

\begin{abstract}
Commercial FPGAs, such as AMD Versal devices, increasingly incorporate AI engines that exploit low-precision packed-SIMD fused multiply–accumulate (FMA) to achieve proportional throughput gains. However, trans-precision FMA (e.g., multiplying two FP16 numbers and adding their result to an FP32 accumulator), which preserves numerical stability by accumulating in higher precision, remains bottlenecked by the highest-precision, lowest-throughput operation. Dot-product accumulation (DPA) (e.g., performing a dot-product on two 4-element FP8 vectors and adding its result to an FP32 accumulator) can fully utilize the input/output bandwidth and computational resources. Existing flexible open-source FPUs, such as FPnew, do not support DPA and implement SIMD FMA on low-precision formats by replicating independent FMA lanes, which increases area, underutilizes shared arithmetic resources, and complicates the integration of DPA operations.

This paper presents TransDot, a reconfigurable FPU that unifies multi-precision SIMD FMA and trans-precision DPA within a shared, reconfigurable datapath. TransDot extends the baseline design with 2-term FP16, 4-term FP8, and 8-term FP4 dot-product accumulation into FP32 using reconfigurable subcomponents. Evaluation shows that TransDot delivers 2$\times$ FP16, 4$\times$ FP8, and 8$\times$ FP4 throughput via DPA with FP32 accumulation, and 1.46$\times$ area efficiency in FP16 DPA and 2.92$\times$ area efficiency in FP8 DPA, at the cost of 37.3\% larger area on average and an additional pipeline stage in dot-product mode compared to the FPnew baseline. These results demonstrate that TransDot’s area-efficient design enables scalable deployment in next-generation AMD Versal AI engines.
\end{abstract}

\begin{IEEEkeywords}
Floating-point unit, dot-product accumulation, mixed-precision arithmetic, FPGA AI engine, reconfigurable datapath
\end{IEEEkeywords}

%

\section{Introduction}
Modern AI accelerators increasingly rely on reduced-precision arithmetic to improve performance and energy efficiency.
Low-precision formats such as FP16~\cite{fp16}, bfloat16, FP8~\cite{fp8_1,fp8_2}, and FP4~\cite{nvfp4} have become widely adopted in both training and inference, with computation dominated by repeated multiply–accumulate operations~\cite{tpu,reduce_precision}. However, accumulation needs to be performed at higher precision to preserve numerical stability, as each output accumulates many products and low-precision accumulation leads to excessive rounding error and degraded convergence~\cite{mixed_precision,lut_tensor}. Trans-precision FMA execution, which involves low-precision multiplication with higher-precision accumulation, has been widely adopted in contemporary AI accelerators such as NVIDIA Tensor Cores~\cite{blackwell}, Google TPU~\cite{tpu_v2}, and AWS Trainium~\cite{trainium}. 

\begin{figure}[tp!]
    \centering
    \includegraphics[width=\linewidth]{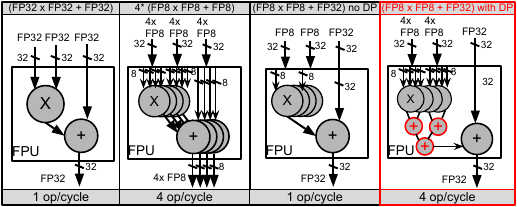}
    \caption{Throughput comparison among FP32 scalar FMA, FP8 packed SIMD FMA, and FP8-to-FP32 trans-precision FMA, with and without native dot-product accumulation (DPA) support.}
    \label{fig:compare}
\end{figure}

\begin{figure}[tp!]
    \centering
    \includegraphics[width=\linewidth]{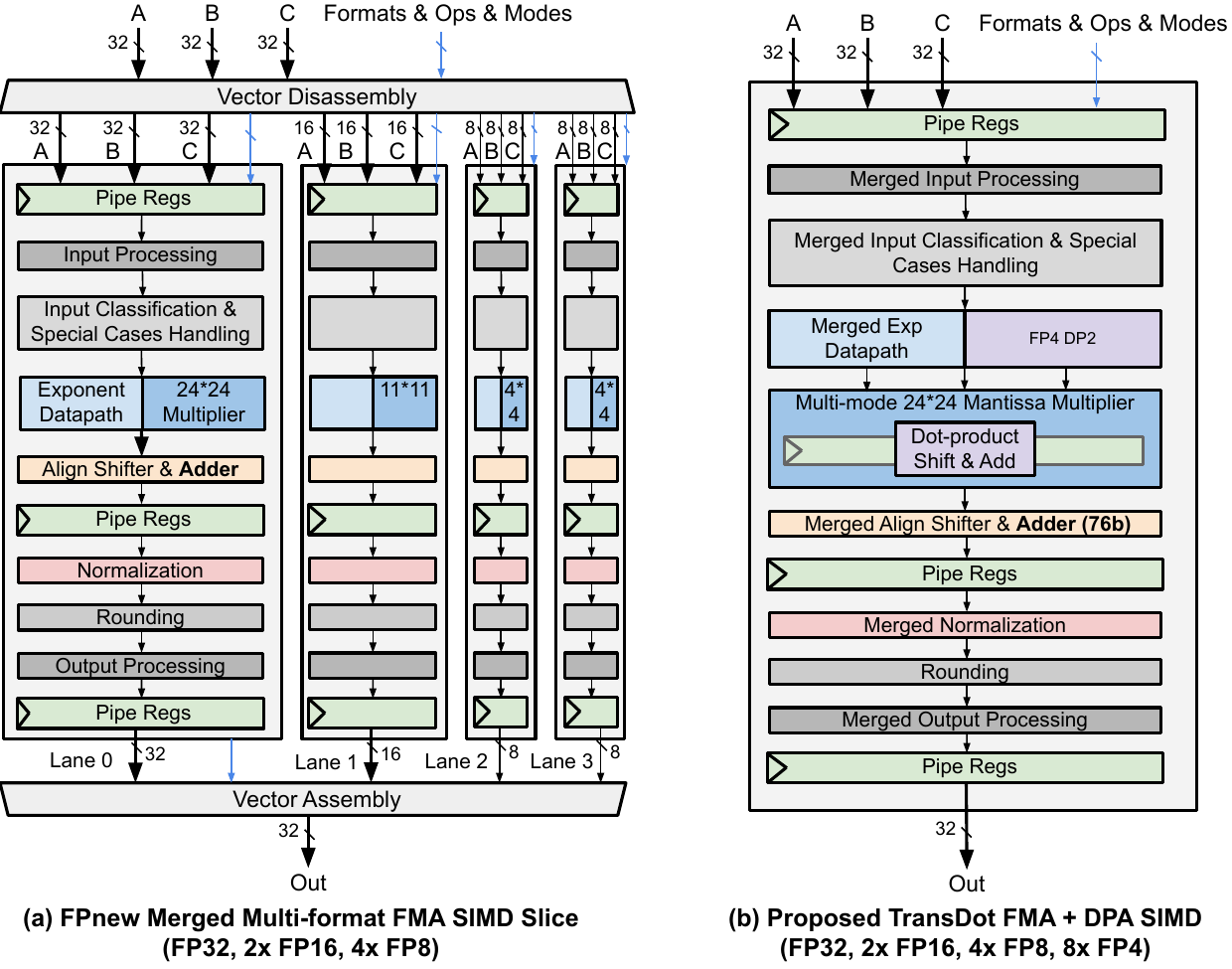}
    \caption{FPnew FMA SIMD slice and TransDot microarchitecture.}
    \label{fig:fpnew}
    \vspace{-1em}
\end{figure}

Commercial FPGAs such as AMD Versal~\cite{amd_versal_ai_engine} increasingly integrate reconfigurable AI engines that exploit low-precision packed-SIMD FMA for proportional throughput gains. However, supporting trans-precision FMA directly in SIMD form would require widening FPU output interfaces to produce multiple high-precision results per cycle, significantly increasing buffer bandwidth and exacerbating input–output imbalance. In practice, FPU input and output interfaces are typically fixed-width, which can produce only a single high-precision result per cycle when software workloads require trans-precision execution, forcing SIMD datapaths and input bandwidth to be underutilized, as illustrated in Fig.~\ref{fig:compare}. Dot-product accumulation (DPA) execution overcomes this limitation by packing reduced-precision operands into a SIMD datapath and accumulating all products into a higher-precision output. 
As Fig.~\ref{fig:compare} shows, by collapsing multiple low-precision products into one high-precision result, DPA preserves FPU input and output bandwidth while achieving throughput comparable to non-trans-precision SIMD execution within a single FPU.

\begin{table}[t!]
\centering
\caption{TransDot Supported Precision Modes for FMA and Dot-Product Accumulation}
\label{tab:transdot_modes}
\setlength{\tabcolsep}{3pt} 
\begin{tabular*}{\columnwidth}{@{\extracolsep{\fill}}lcccc}
\toprule
\textbf{Format} & \textbf{Encoding} & \textbf{Scalar/SIMD FMA} & \textbf{DPA} & \textbf{Accumulate Format} \\
\midrule
FP32 & E8M23  & 1-way & 1-term & FP32 \\
FP16 & E5M10  & 2-way & 2-term & FP32 / FP16 \\
FP8  & E4M3 & 4-way & 4-term & FP32 / FP16 \\
FP4  & E2M1 & 8-way & 8-term & FP32 / FP16 \\
\bottomrule
\end{tabular*}
\vspace{-1em}
\end{table}

\begin{figure}[t]
    \centering
    \includegraphics[width=\linewidth]{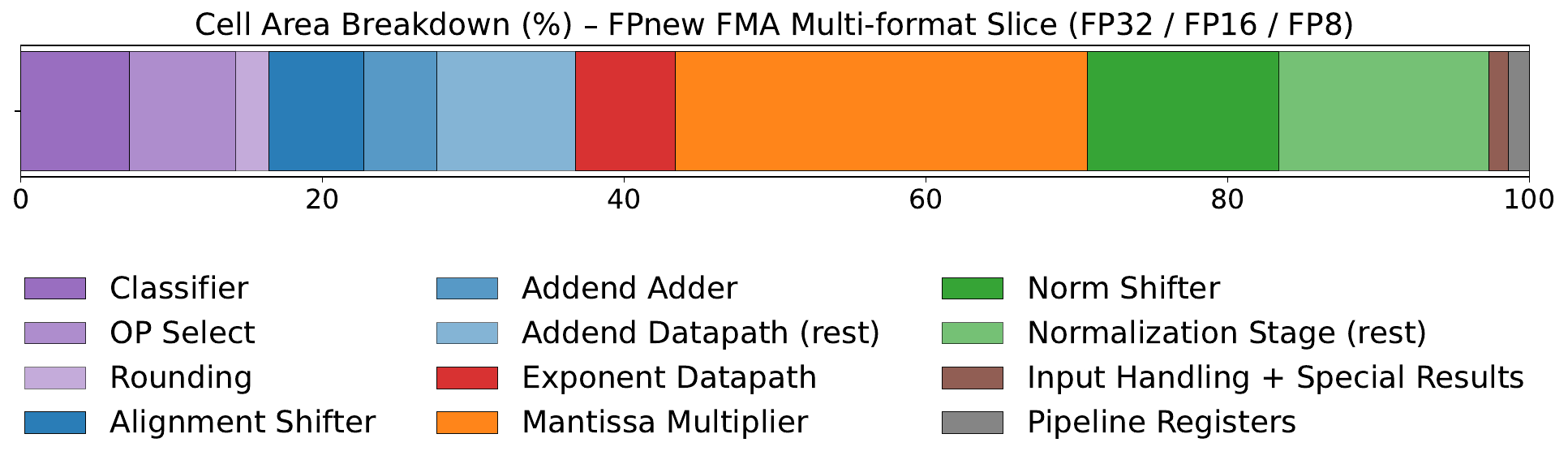}
    \caption{Area breakdown (\%) for an FPnew multi-format FMA slice.}
    \label{fig:area_percentage}
\end{figure}
Existing open-source FPUs such as FPnew~\cite{fpnew} and HardFloat~\cite{hardfloat} provide flexible format configurability and support packed-SIMD reduced-precision operations, but they lack native support for trans-precision DPA. Several prior works~\cite{dpu_1,dpu_2,fusion-3d} have explored dedicated DPA units, such as Minifloats~\cite{dpu_0}, which designed a dot-product unit to be integrated into FPnew~\cite{fpnew} as a separate compute slice. However, these units can become underutilized when workloads expose limited dot-product parallelism, while their dedicated datapaths restrict reuse across other execution modes. Other works~\cite{fma_0,fma_1} propose multi-format FMA units with mixed-precision dot-product support for higher-precision formats, but they predate emerging AI datatypes such as FP8 and FP4, and~\cite{fma_0} is limited to two-term dot products.

TransDot bridges these design points by extending a scalar FP32 unit to support both multi-format SIMD FMA and trans-precision DPA through systematic datapath reuse, including modern AI formats. Unlike Minifloats~\cite{dpu_0}, which adds a dedicated dot-product unit alongside FPnew's FMA slices, TransDot reuses the existing FP32 datapath to enable dot-product execution with minimal architectural modifications, significantly reducing area overhead and making it well-suited for integration into FPGA AI engines such as Versal AI Engines~\cite{amd_versal_ai_engine}. This paper makes the following contributions:
\begin{itemize}
  \item We present \textbf{TransDot}, a reconfigurable FPU that supports trans-precision DPA and unifies scalar, SIMD, and DPA execution within a largely shared datapath.
  \item We extend the supported format set to include emerging low-precision floating-point formats, including FP4, as summarized in Table~\ref{tab:transdot_modes}.
  \item We evaluate TransDot using ASIC flow with a commercial 12nm technology, demonstrating favorable area–delay trade-offs and strong area efficiency.
\end{itemize}

\section{Architecture and Methodology}
\subsection{FPnew (Baseline) Floating-Point FMA Datapath}
Fig.~\ref{fig:fpnew}(a) illustrates the open-source FPnew multi-format FMA slice, which supports FP32, FP16, and FP8 execution in both scalar and SIMD modes. The FMA ($A \times B + C$) is implemented by multiplying the mantissas of $A$ and $B$ to generate a carry-save product, while the exponent datapath computes the alignment shift for operand $C$. An alignment shifter aligns the mantissa of $C$ with the product, after which the aligned operands are accumulated using an adder sized to a no-precision-loss range ($(3p+4)$, where $p$ denotes the precision in bits, i.e., mantissa\_bits + 1).

FPnew enables SIMD execution by instantiating additional narrow lanes for lower-precision formats, while a shared wide datapath supports multi-format mixed-precision FMA. Although this organization is clean and modular, it leads to suboptimal hardware utilization both in scalar mode and SIMD mode. Moreover, FPnew does not natively support trans-precision DPA. As a result, when accumulating into FP32, FPnew can issue only a single FMA per cycle, limiting throughput compared to reduced-precision SIMD execution.

In contrast, TransDot introduces explicit DPA support for various formats, allowing trans-precision FMA to achieve SIMD-equivalent throughput. For better area efficiency, TransDot abandons FPnew’s lane-replication approach in favor of fine-grained datapath reuse across precision modes through configurable arithmetic and logic subcomponents. The microarchitecture of TransDot is shown in Fig.~\ref{fig:fpnew}(b).

\subsection{Subcomponent Datapath Reuse}
In TransDot, wide datapath elements are reused as shared substrates and reconfigured through partitioning, gating, and mode-dependent control to support scalar, SIMD, and DPA execution. Components that operate independently at the bit level, such as logic operators, are reused directly, while hierarchical arithmetic structures (e.g., shifters, adders, and multipliers) are restructured to expose internal subword parallelism. Format-specific stages, such as rounding and special case handling, are duplicated for each SIMD lane as lightweight peripheral logic. Fig.~\ref{fig:area_percentage} presents an area breakdown of an FPnew FMA slice, showing that shifters and multipliers dominate overall area. Accordingly, the following subsections focus on these components and describe how TransDot transforms them into reconfigurable, multi-mode structures that efficiently support trans-precision DPA.

\subsubsection{Reconfigurable Barrel Shifter}
\begin{figure}[tp!]
    \centering
    \includegraphics[width=0.7\linewidth]{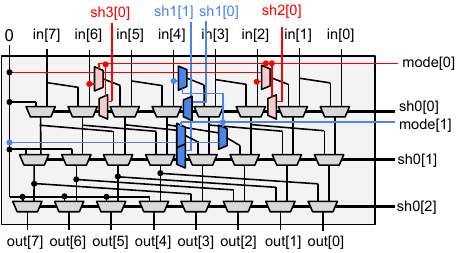}
    \caption{Reconfigurable barrel shifter architecture. Blue regions indicate additional hardware enabling two half-width modes, while red regions show further extensions supporting four quarter-width modes. sh: shift amount signals; In quarter mode, mode[1:0]=2'b11; in half mode, mode[1:0]=2'b10}
    \label{fig:shifter}
\end{figure}

Shifters are used multiple times within an FMA datapath. The alignment stage employs a right shifter of width $(4p+4)$ bits (100 bits for FP32), while the normalization stage requires a left shifter of width $(3p+5)$ bits. Together, these large shifters account for approximately 15-20\% of the total FPU area (Fig.~\ref{fig:area_percentage}). To reduce this overhead across multiple execution modes, TransDot introduces a multi-mode shifter that supports full-width, two half-width, and four quarter-width operations, built upon a conventional barrel shifter architecture.

Fig.~\ref{fig:shifter} illustrates the microarchitecture of the proposed reconfigurable barrel shifter using an 8-bit example for clarity. The area overhead of supporting multiple modes can be estimated by the additional multiplexers required. A conventional $n$-bit barrel shifter consists of $\log_2(n)$ stages, each containing $n$ 2:1 multiplexers, for a total of $n \log_2(n)$ multiplexers.

To enable reconfiguration, additional logic is required to handle mode-dependent shift amounts, prevent data from crossing subword boundaries, and selectively bypass the final stages when operating on narrower partitions. The total number of additional multiplexers introduced by these mechanisms can be calculated as
$
\frac{5n}{8} + 3\log_2(n) - 5.
$

Based on this estimate, the reconfigurable barrel shifter incurs an area overhead of approximately 10.7\% for $n=128$ and 13.8\% for $n=64$, relative to a baseline barrel shifter. This estimate excludes high-fanout buffering, which increases the baseline area and further amortizes the relative overhead in practice. In contrast, implementing four fully independent shifters would incur substantially higher overheads of approximately 78.5\% for $n=128$ and 75\% for $n=64$, highlighting the efficiency of the proposed reconfigurable design.

\begin{figure}[tp!]
    \centering
    \includegraphics[width=0.9\linewidth]{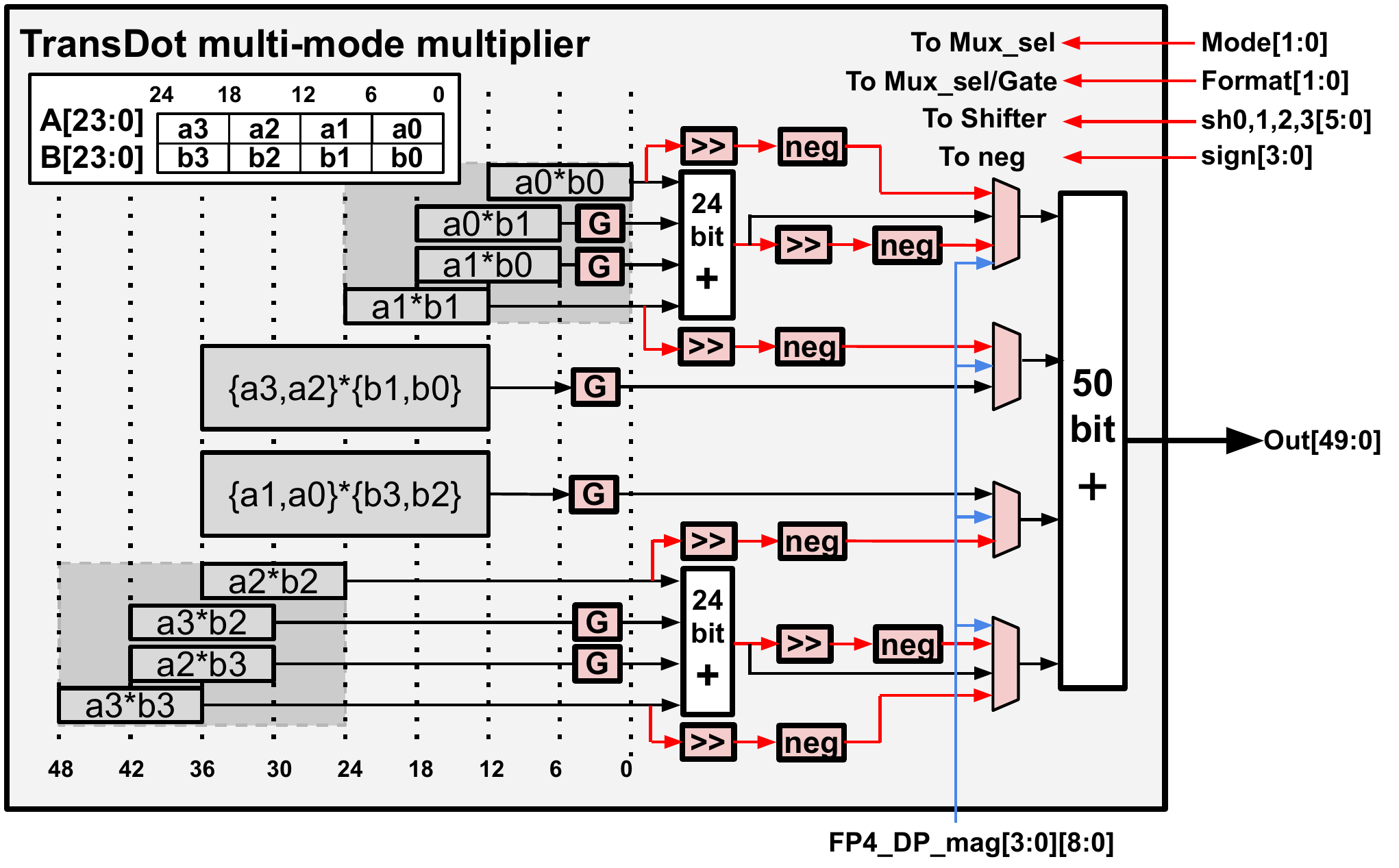}
    \caption{TransDot multi-mode multiplier architecture. The red part shows new hardware for dot-product operations compared to an array multiplier. (G: Gate).}
    \label{fig:array_multiplier}
\end{figure}

\subsubsection{Multi-mode Array Multiplier}
The mantissa multiplier is a major contributor to area (about 30\% in Fig.~\ref{fig:area_percentage}) in the FMA datapath and is considerably harder to reuse across precision modes than adders or logic units. A conventional $p$-bit mantissa multiplier is typically implemented as a monolithic structure, resulting in substantial underutilization when executing reduced-precision or SIMD operations. Prior work has explored reusing internal partial products in array or Booth multipliers to support lower precisions, allowing a $2n$-bit multiplier to be partitioned into two $n$-bit multipliers and so on~\cite{simd_only,fma_0,booth}. However, as the operand width decreases, large portions of the multiplier, including sub-multipliers and partial-sum adders, remain idle. To improve utilization,~\cite{simd_only} repurposes these idle sub-multipliers to generate additional SIMD results. We further observe that the reduced adder tree exposed in such decomposed multipliers naturally matches the accumulation pattern of dot-product operations, presenting an opportunity for deeper datapath reuse beyond SIMD execution.

Fig.~\ref{fig:array_multiplier} illustrates the proposed multi-mode multiplier, which enables datapath reuse across scalar, SIMD, and DPA modes. The original 24-bit mantissas in FP32 mode are partitioned into four 6-bit segments. 8 12-bit and 2 24-bit partial products are generated once and selectively combined according to the execution mode. In scalar mode, the full adder tree is enabled to produce a full-precision result. In low-precision SIMD and DPA mode, corresponding partial sums are gated. In DPA mode, six shifters are added for alignment shifting, and six negate units are added to determine conditional negation for subtraction cases. A reconfigurable pipeline stage can be inserted to alleviate timing pressure across modes.

\subsubsection{FP4 Support}
Due to the extremely limited dynamic range of FP4 operands, TransDot employs a dedicated FP4 2-term dot-product (DP2) stage that directly computes the products of two FP4 operand pairs in sign–magnitude form. This stage generates four 9-bit partial products along with their corresponding signs from 8 pairs of FP4 inputs, which are then forwarded to the multi-mode multiplier for final accumulation into a shared higher-precision result.

\section{Evaluation and Results}
We evaluate TransDot through RTL modification and ASIC synthesis using Cadence Genus~\cite{cadence_genus} with a commercial 12nm PDK. A key concern in this evaluation is ensuring that the proposed RTL changes are not eliminated or obscured by aggressive synthesis optimizations, or that equivalent optimizations are not automatically inferred from a less optimized RTL. Accordingly, we configured Genus with high mapping and optimization effort and enabled retiming with high effort to enforce maximal area recovery under the specified timing constraints. This setup ensures that any observed area differences reflect deliberate microarchitectural changes rather than synthesis artifacts. We begin our evaluation by analyzing area savings at the subcomponent level.

\subsection{Reconfigurable Barrel Shifter}

\begin{figure}[tp!]
  \centering

  \begin{subfigure}[t]{0.48\linewidth}
    \centering
    \includegraphics[width=\linewidth]{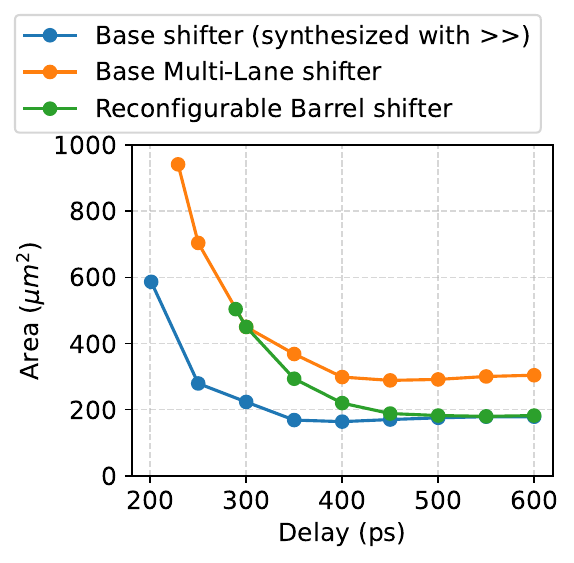}
    \caption{Reconfigurable shifters.}
    \label{fig:shifter_area_delay}
  \end{subfigure}
  \hfill
  \begin{subfigure}[t]{0.48\linewidth}
    \centering
    \includegraphics[width=\linewidth]{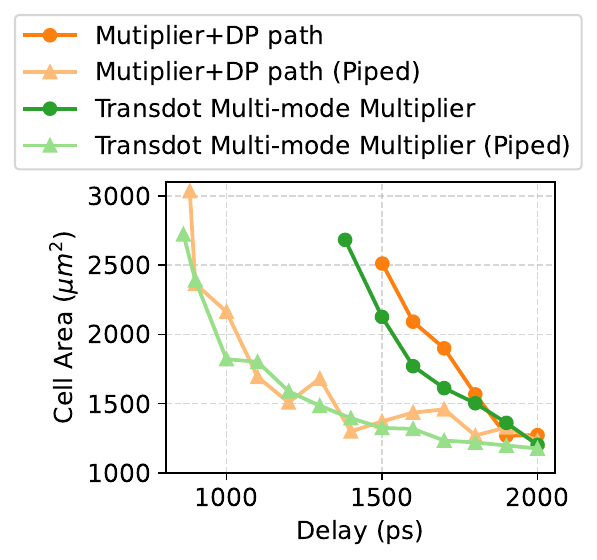}
    \caption{Multi-mode array multipliers.}
    \label{fig:multiplier_area_delay}
  \end{subfigure}

  \caption{Area–delay trade-offs of TransDot reconfigurable datapath components.
  (a) Area vs. delay for a 100-bit single-mode shifter, a 100-bit multi-lane shifter with four lanes (FPnew-style: full-width + half-width + two quarter-width shifters), and the proposed 100-bit reconfigurable barrel shifter.
  (b) Area vs. delay comparing a multiplier with a separated dot-product datapath against the TransDot multi-mode array multiplier (pipelined vs. unpipelined).}
  \label{fig:area_delay_components}
  \vspace{-1em}
\end{figure}

Fig.~\ref{fig:shifter_area_delay} compares the synthesized area of three shifter designs (described in the caption) in Genus. 
When the target delay exceeds 400~ps, the area of the reconfigurable shifter converges to that of the baseline synthesized shifter, while the multi-lane shifter remains 35.8\% to 67.2\% larger. As the delay constraint is tightened below 400ps, the area of the reconfigurable shifter increases and approaches that of the multi-lane design, indicating that the synthesis tool progressively replicates or decouples datapath segments to trade area for improved performance. These results show that the proposed barrel shifter can achieve significantly lower area through explicit datapath sharing, capturing optimization opportunities that are difficult for synthesis tools to infer automatically.

\subsection{TransDot Multi-mode Multiplier}
Fig.~\ref{fig:multiplier_area_delay} shows the cell-area versus delay trade-off of separated multipliers with a dedicated dot-product reduction path and the proposed TransDot multi-mode multiplier. The results show that TransDot achieves both lower area and better timing than the separated baseline in the combinational case: the minimum effective delay of TransDot is about 1.38 ns, whereas the separated design achieves about 1.50 ns, and TransDot reduces cell area by 15.4\% at 1.6 ns. This improvement comes from reusing the adder structure inside the unified datapath, which shortens the critical reduction path while avoiding dedicated dot-product logic. With a single inserted pipeline stage, the minimum effective delay is about 0.86 ns for TransDot and 0.88 ns for the separated design. At 1.0 ns, the pipelined TransDot design reduces cell area by 15.8\%, and it continues to provide lower area at more relaxed timing points. Overall, these results show that arithmetic reuse in TransDot improves area efficiency without sacrificing frequency.

\subsection{TransDot Performance, Area, and Energy}
\begin{figure}[tp!]
  \centering
  \begin{subfigure}[t]{0.48\linewidth}
    \centering
    \includegraphics[width=\linewidth]{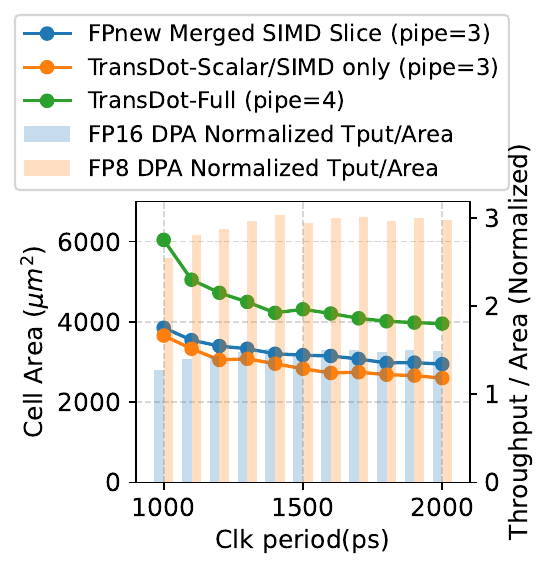}
    \caption{Area–delay and normalized throughput/area (TransDot/FPnew).}
    \label{fig:top_result}
  \end{subfigure}
  \hfill
  \begin{subfigure}[t]{0.48\linewidth}
    \centering
    \includegraphics[width=\linewidth]{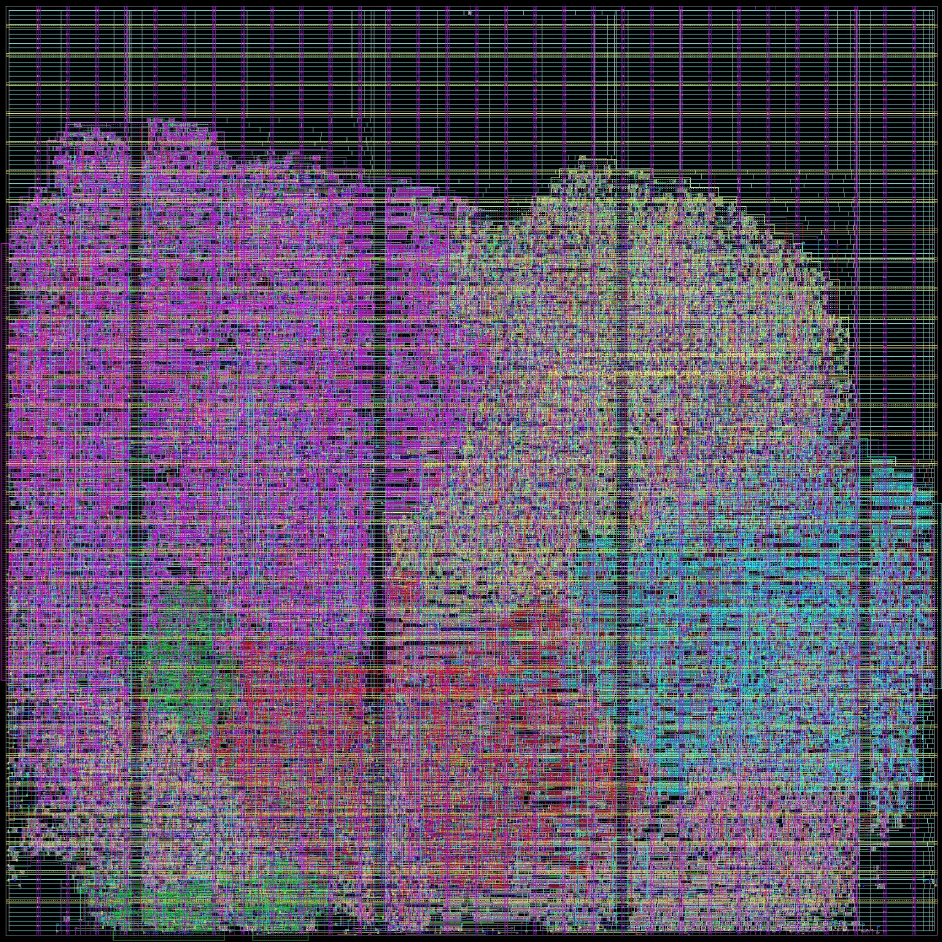}
    \caption{TransDot layout (100$\mu$m $\times$ 100$\mu$m, 1 GHz)}
    \label{fig:layout}
  \end{subfigure}

  \caption{Area–delay comparison and physical layout of TransDot. 
  Pink: multi-mode multiplier (34.5\%); cyan: normalization (15.5\%); red: exponent (11.8\%); yellow: alignment shifter and adder (18.1\%); green: FP4 DP2 (3.9\%); purple: others (16.2\%).}
  \label{fig:area_delay_layout}
\end{figure}

Fig.~\ref{fig:top_result} compares the area–delay trade-offs of the FPnew baseline, TransDot with merged SIMD lanes, and the full TransDot, along with the relative area efficiency (measured in throughput/area). Fig.~\ref{fig:layout} shows the corresponding physical layout generated using Synopsys ICC2~\cite{icc2}. The FP4-related logic only adds 3.9\% extra overhead. TransDot with merged SIMD lanes achieves 9.44\% area reduction relative to FPnew on average. Extending the design to the full TransDot configuration enables 2$\times$ FP16, 4$\times$ FP8, and 8$\times$ FP4 FMA throughput via DPA, while increasing area by 37.3\% (31.8\% to 56.8\%) on average over the baseline. As a result, TransDot achieves 1.46$\times$ (1.28$\times$ to 1.52$\times$) area efficiency for FP16 DPA and 2.92$\times$ (1.56$\times$ to 3.04$\times$) for FP8 DPA relative to the FPnew baseline on average. Table~\ref{tab:transdot_energy} reports the corresponding energy per FLOP across supported precision modes after place-and-route with PrimeTime PX~\cite{primetime_px}. 

\begin{table}[t!]
  \centering
  \caption{Performance and energy efficiency of TransDot (Post-pnr, 12 nm, 1 GHz, 0.8 V nominal, TT-corner).}
  \label{tab:transdot_energy}
  \begin{threeparttable}
  \begin{tabular}{lccc}
    \toprule
    \textbf{Precision/operation} & \textbf{Lat/T*} & \textbf{Perf} & \textbf{Energy} \\
                                  &  & \textbf{(GFLOP/s)} & \textbf{(pJ/FLOP)}\\
    \midrule
    FP32 FMA Scalar & 4/1 & 2 & 3.75 \\
    FP16 FMA Scalar& 4/1 & 2 & 2.76\\
    FP16 FMA SIMD & 4/1 & 4 & 1.85 \\
    FP16 DPA with FP32 Acc & 4/1 & 4 & 1.80  \\
    FP8 FMA Scalar  & 4/1 & 2 & 2.21 \\
    FP8 FMA SIMD  & 4/1 & 8 & 0.84 \\
    FP8 DPA with FP32 Acc & 4/1 & 8 & 0.84  \\
    FP4 DPA with FP32 Acc & 4/1 & 16 & 0.41 \\
    \bottomrule
  \end{tabular}
  \begin{tablenotes}[flushleft]
  \footnotesize
  \item[*] Latency (cycles) / Throughput (ops/cycle)
  \end{tablenotes}
  \end{threeparttable}
  \vspace{-1pt}
\end{table}

\section{Conclusion}
This paper presents TransDot, a reconfigurable FPU that unifies SIMD FMA and trans-precision DPA within a shared reconfigurable datapath. 
Evaluation shows that TransDot 
achieves 1.46$\times$ area efficiency for FP16 DPA and 2.92$\times$ area efficiency for FP8 DPA at the cost of 37.3\% extra area on average compared with the baseline. These results demonstrate that TransDot is a practical and area-efficient building block for next-generation AMD Versal AI Engines. 
The RTL implementation of TransDot is publicly available at: \url{https://github.com/pncel/TransDot}.

\newpage



%



\bibliographystyle{IEEEtran}
\bibliography{refs}

\end{document}